\documentclass[aps,preprint,10pt,twocolumn]{revtex4}%
\usepackage{amsfonts}
\usepackage{amsmath}
\usepackage{amssymb}
\usepackage[pagebackref]{hyperref}
\usepackage{graphicx}%
\providecommand{\U}[1]{\protect\rule{.1in}{.1in}}

\begin{document}
\preprint{ }
\title{$SL(2,%
\mathbb{R}
)$ matrix model and supersymmetric Yang-Mills integrals}
\author{Miguel Tierz}
\affiliation{Institut d'Estudis Espacials de Catalunya (IEEC/CSIC). Campus UAB, Facultat de
Ci\'{e}ncies, Torre C5-Parell-2a planta. E-08193 Bellaterra (Barcelona) Spain.}
\homepage{http://www.tierz.com/}
\email{tierz@ieec.fcr.es}

\begin{abstract}
The density of states of Yang-Mills integrals in the supersymmetric case is
characterized by power-law tails whose decay is independent of $N$, the rank
of the gauge group. It is believed that this has no counterpart in matrix
models, but we construct a matrix model that exactly exhibits this property.
In addition, we show that the eigenfunctions employed to construct the matrix
model are invariant under the collinear subgroup of conformal transformations,
$SL(2,%
\mathbb{R}
).$ We also show that the matrix model itself is invariant under a fractional
linear transformation. The wave functions of the model appear in the
trigonometric Rosen-Morse potential and in free relativistic motion on $AdS$ space.

\end{abstract}
\maketitle

\section{Introduction}

The dimensional reduction of ten-dimensional Yang-Mills theory to $p+1$
dimensions has drawn interest over a considerable period of time, due to its
relevance to the description of $p$-dimensional $D$-branes in string theory.
The functional integral of Yang-Mills theory becomes an ordinary
multi-dimensional integral, which is well-defined and finite if supersymmetry
is present \cite{Krauth:1998xh,Krauth:1998yu,Krauth:1999qw}. Explicitly, the
integral reads, for gauge group $SU(N)$,%

\begin{align}
\mathcal{Z}_{D,N}  &  :=\int\prod_{A=1}^{N^{2}-1}\left(  \prod_{\mu=1}%
^{D}\frac{dX_{\mu}^{A}}{\sqrt{2\pi}}\right)  \left(  \prod_{\alpha
=1}^{\mathcal{N}}d\Psi_{\alpha}^{A}\right)  \times\label{susyint}\\
&  \exp\left[  \frac{1}{2}\mathrm{Tr}\,[X_{\mu},X_{\nu}][X_{\mu},X_{\nu
}]+\mathrm{Tr}\,\Psi_{\alpha}[\Gamma_{\alpha\beta}^{\mu}X_{\mu},\Psi_{\beta
}]\right]  .\nonumber
\end{align}
The matrices in the exponent in eq.(\ref{susyint}) are in the fundamental
representation of $SU(N)$, i.e. $X_{\mu}=X_{\mu}^{A}T_{A}$, $\Psi_{\alpha
}=\Psi_{\alpha}^{A}T_{A}$, where the $SU(N)$ generators $T_{A}$ are Hermitian
and normalized such that $\mathrm{Tr}\,T^{A}T^{B}=\frac{1}{2}\delta^{AB}$. The
symmetric $\mathcal{N}\times\mathcal{N}$ matrices $\Gamma^{\mu}$ are related
to the standard $SO(1,D-1)$ gamma matrices by $\Gamma^{\mu}=\mathcal{C}%
\gamma^{\mu}$, where $\mathcal{C}$ is the charge conjugation matrix. The model
is supersymmetric in dimensions $D=3,4,6,10$, where the degree $\mathcal{N}$
of (real) supersymmetry is, respectively, $\mathcal{N}=2(D-2)=2,4,8,16$, with
the supersymmetry variations
\begin{equation}
\delta X_{\mu}=i\bar{\varepsilon}\gamma^{\mu}\Psi\qquad\delta\Psi=-\frac{i}%
{2}[X_{\mu},X_{\nu}]\gamma^{\mu}\gamma^{\nu}\varepsilon.
\end{equation}
Recall that the case $D=10$ is essentially the IKKT model of the
$\mathrm{IIB}$ superstring \cite{Ishibashi:1996xs}. Besides, since the
reduction is to zero dimensions ($p=-1$), we are thus describing the
configuration space of N $D$-instantons \cite{Green:1997tn}. For a general
introduction to Yang-Mills integrals see the thesis \cite{Austing:2001ib}, for example.

In this letter, we shall mainly focus on the behavior of the density of states
of Yang-Mills integrals in the supersymmetric case, that shows a sharp
contrast with the more standard behavior of the bosonic case. The distribution
of eigenvalues for the eigenvalues $\lambda_{i}$ of just one matrix, say,
$X_{1}$, in the background of the other matrices $X_{2},\ldots,X_{D}:$
\begin{equation}
\rho(\lambda)={\frac{1}{N}}\bigg\langle\sum_{i=1}^{N}\delta(\lambda
-\lambda_{i})\bigg\rangle, \label{density}%
\end{equation}
Here, the average $<>$ is with respect to eq.(\ref{susyint}). In particular,
note the following result/remark in \cite{Krauth:1999qw}:

\emph{"Most strikingly, the decay of the densities in the susy cases
}$D=4,6,10$\emph{\ (}$\rho(\lambda)\sim\lambda^{-3},\lambda^{-7},\lambda
^{-15}$\emph{) is independent of }$N$\emph{. It means that the eigenvalue
distribution are wide even in the }$N\rightarrow\infty$\emph{\ limit! This is
a most unusual effect for a random matrix model. Evidently, supersymmetry is
responsible for this behavior."}

In the following, we shall show that one can naturally construct an ordinary
random matrix model with this very same property. We shall construct it using
orthogonal polynomials. Then, we shall address the significance and uniqueness
of the particular choice of matrix model. Let us consider a generic Hermitian
matrix model:%

\begin{equation}
Z=C_{N}\int\prod_{i=1}^{N}\mathrm{e}^{-V(x)}\prod_{i<j}\left(  x_{i}%
-x_{j}\right)  ^{2}dx_{i}, \label{sinh}%
\end{equation}
This model has a density states that can be exactly computed with the
polynomials orthogonal w.r.t. $\mathrm{e}^{-V(x)}$:%

\begin{equation}
\rho\left(  x\right)  =\sum_{n=0}^{N-1}\phi^{2}\left(  x\right)  ,\text{ }%
\phi\left(  x\right)  =\mathrm{e}^{-V(x)}P_{n}\left(  x\right)  \label{dens}%
\end{equation}%
\[
\int\mathrm{e}^{-V(x)}P_{n}\left(  x\right)  P_{m}\left(  x\right)
dx=h_{n}\delta_{n,m}.
\]
Now, we can construct a matrix model using the wavefunctions:
\begin{equation}
\phi\left(  x\right)  =\left(  1+\lambda x^{2}\right)  ^{-N-\alpha}%
P_{n}\left(  x\right)  , \label{waves}%
\end{equation}
The Cauchy case, $\alpha=0$ is easy to compute exactly and gives:%
\begin{equation}
\rho\left(  x\right)  =\frac{1}{1+\lambda x^{2}},\text{ }\forall N
\end{equation}
The reason to have exactly the same expression for any $N$ is due to the fact
that the weight function (potential) changes with $N$ as well.

Note that to match the Yang-Mills density of states, the full computation of
$\left(  \ref{dens}\right)  $ with $\left(  \ref{waves}\right)  $ is not
necessary. It is easy to directly see that one obtains power-law tails whose
decay is independent of $N$. The weight function part is multiplied by a
polynomial whose highest order is $x^{2N-2}$, hence:%

\begin{align}
\rho\left(  x\right)   &  =\frac{1}{\left(  1+x^{2}\right)  ^{N+\alpha}}%
\sum_{n=1}^{N-1}P_{n}^{2}\left(  x\right)  \Rightarrow\rho\left(  x\right)
\sim\frac{x^{2N-2}}{x^{2\left(  N+\alpha\right)  }}\\
&  =\frac{1}{x^{2(\alpha+1)}}\text{ for }x\rightarrow\infty.\nonumber
\end{align}
The power-law behavior of susy Yang-Mills integrals and the corresponding
$\alpha$ parameter:%

\begin{equation}%
\begin{tabular}
[c]{|c|c|c|}\hline
$D=4$ & $\rho\left(  x\right)  \sim x^{-3}$ & $\alpha=\frac{1}{2}$\\\hline
$D=6$ & $\rho\left(  x\right)  \sim x^{-7}$ & $\alpha=\frac{5}{2}$\\\hline
$D=10$ & $\rho\left(  x\right)  \sim x^{-15}$ & $\alpha=\frac{13}{2}$\\\hline
\end{tabular}
\end{equation}
Before proceeding to discuss properties of this random matrix model and its
relationship with more standard ones, let us note that the correspondence with
the Yang-Mills susy behavior holds not only for large $N$, but also for all
$N$. Besides, the independence with the dimension $N$, which is the truly
unusual feature for a matrix model density of states, is due to the fact that
one actually has a different weight function with each $N$. The increased
eigenvalue repulsion is exactly compensated by an increase in the confining
properties of the potential. In addition, notice the following property
satisfied by the weight $\omega\left(  x\right)  :$%

\begin{align}
x  &  \rightarrow x^{\prime}=\frac{ax+b}{cx+d},\\
\omega\left(  x\right)   &  \rightarrow\omega^{\prime}\left(  x\right)
=\left(  cx+d\right)  ^{-2j}\omega\left(  \frac{ax+b}{cx+d}\right) \nonumber
\end{align}
which is the defining property of a function being invariant under $SL(2,%
\mathbb{R}
),$ the "collinear" subgroup of the conformal group $SO(2,4)$ \cite{Lang}.
Thus, the weight function is invariant under $SL(2,\mathrm{%
\mathbb{R}
})$. In our models $j=-\alpha$.

According to \cite{Witte}, the weight in $\left(  \ref{waves}\right)  $ should
be considered "classical", on equal footing with the Gaussian (Hermite),
Laguerre and Jacobi cases. However, the definition of a classical random
matrix \cite{classical} is actually satisfied by a much larger set of models,
that can be characterized by a generic weight function:%
\begin{equation}
\omega\left(  x\right)  =\exp\left(
{\displaystyle\int}
\frac{\left(  d-2a\right)  x+\left(  e-b\right)  }{ax^{2}+bx+c}dx,\right)
\label{Pearson}%
\end{equation}
with $\sigma\left(  x\right)  =ax^{2}+bx+c$ and $\tau\left(  x\right)  =dx+e$
the polynomials in the differential equation:%
\begin{equation}
\sigma\left(  x\right)  P_{n}^{\prime\prime}(x)+\tau\left(  x\right)
P_{n}^{\prime}\left(  x\right)  +P_{n}\left(  x\right)  =0
\end{equation}
satisfied by the polynomials $P_{n}\left(  x\right)  $ orthogonal to the
weight $\left(  \ref{Pearson}\right)  $ \cite{Tierz}. If we consider
$a=1,b=0,c=1,d=2\left(  1-\eta\right)  ,e=0$ in $\left(  \ref{Pearson}\right)
$ we have then:%
\begin{equation}
\omega\left(  x\right)  =\frac{1}{\left(  1+x^{2}\right)  ^{\eta}}\text{ ,}%
\end{equation}

\begin{equation}
\left(  1+x^{2}\right)  P_{n}^{\prime\prime}(x)+2\left(  1-\eta\right)
xP_{n}^{\prime}\left(  x\right)  +P_{n}\left(  x\right)  =0 \label{diffeq}%
\end{equation}
We have specified as few parameters as possible. The model can be considered
in a much more general form with the choice \cite{KM}:%
\begin{align}
a  &  =A^{2}+C^{2},\text{ }b=2(AB+CD)\text{, }c=B^{2}+D^{2}\\
d  &  =2(1-\eta)\left(  A^{2}+C^{2}\right)  \text{ }\nonumber\\
e  &  =q\left(  AD-BC\right)  +2(1-\eta)\left(  AB+CD\right)  ,\nonumber
\end{align}
that leads to a weight function:%

\begin{equation}
\omega\left(  x\right)  =\frac{\exp\left(  q\arctan\left(  \frac{Ax+B}%
{Cx+D}\right)  \right)  }{\left(  \left(  Ax+B\right)  ^{2}+(Cx+D)^{2}\right)
^{\eta}}.
\end{equation}
and makes manifest the invariance under projective transformations. Let us
consider the matrix model:%
\begin{equation}
P(x_{1},...,x_{N})=C_{N}\int\prod_{i=1}^{N}\frac{1}{\left(  1+x_{i}%
^{2}\right)  ^{N+\alpha}}\prod_{i<j}\left(  x_{i}-x_{j}\right)  ^{2}dx_{i},
\end{equation}
and consider the projective transformation of the eigenvalues: $y_{i}%
=\frac{ax_{i}+b}{cx_{i}+d}$, we arrive at:
\begin{align}
P(y_{1},...,y_{N})  &  =C_{N}\int\prod_{i=1}^{N}\frac{\left(  Cy_{i}+D\right)
^{2\alpha}}{\left(  \left(  Ay_{i}+B\right)  ^{2}+\left(  Cy_{i}+D\right)
^{2}\right)  ^{N+\alpha}}\times\label{MM:SL(2,R)}\\
&  \prod_{i<j}\left(  y_{i}-y_{j}\right)  ^{2}dx_{i}.\nonumber
\end{align}
Note how remarkably simple is the Cauchy case ($\alpha=0)$. In particular,
$\left(  \ref{MM:SL(2,R)}\right)  $ generalizes \cite{Brow}, that shows that
the inverse (that corresponds to $a=d=0$ and $b=c=1$ above) of the Cauchy
ensemble is a Cauchy ensemble. Thus, the Hermitian matrix model with the susy
Yang-Mills behavior is characterized by:%
\begin{equation}
P(M)=\exp\left(  -\left(  N+\alpha\right)  \mathrm{Tr}\left(  1+M^{2}\right)
\right)  \label{weight}%
\end{equation}

\begin{align}
M  &  \rightarrow M^{\prime}=\frac{aM+b}{cM+d},\\
P\left(  M\right)   &  \rightarrow P^{\prime}\left(  M\right)  =\left(
cM+d\right)  ^{2\alpha}P\left(  \frac{aM+b}{cM+d}\right)  ,\nonumber
\end{align}
and hence the name $SL(2,%
\mathbb{R}
)$ matrix model.

Systematic exploration of the possibilities in $\left(  \ref{Pearson}\right)
$, shows that the conformal symmetry together with the $N$-independent
power-law tail density of states behavior are a particular feature of this
model. The weight function $\omega\left(  x\right)  =x^{\rho}/(1+x)^{\mu+\rho
}$ , contained in $\left(  \ref{Pearson}\right)  $, leads to power-law tails
that are dimension independent but notice that, in contrast to supersymmetric
Yang-Mills integrals \cite{Krauth:1999qw}, the support is only positive
definite. Let us point out that the matrix model $\left(  \ref{weight}\right)
$ is also very meaningful from the point of view of free probability theory
\cite{Voiculescu}, especially the Cauchy case $\left(  \alpha=0\right)  ,$
that appears in the non-commutative generalization of stable probability
distributions \cite{stable}. But we have seen that the correct match with the
Yang-Mills behavior is obtained for $\alpha\neq0,$ so the $SL(2,%
\mathbb{R}
)$ matrix model should be rather considered a random matrix generalization of
the $t$-Student distribution, which is not a stable but an
infinitely-divisible distribution \cite{Feller}. These mathematical features
of the model will be discussed elsewhere. It is also noteworthy that the
orthogonal polynomials associated to the weight:%
\begin{equation}
\omega\left(  x\right)  =\left(  1+x^{2}\right)  ^{-\eta}\exp\left(  p\arctan
x\right)  ,
\end{equation}
were actually discussed as early as $1929$ \cite{Rom} and, interestingly
enough, they are the solution of a central quantum mechanical problem. Namely,
the Rosen-Morse potential:%
\begin{equation}
V(y)=-2p\cot y+a(a-1)\csc^{2}y,
\end{equation}
which is a well-known model in supersymmetric quantum mechanics and leads to
the same differential equation and orthogonal polynomials employed here. The
case $p=0$ is the P\"{o}schl-Teller potential and corresponds exactly to the
actual set of polynomials of our matrix model $\left(  \ref{diffeq}\right)  $.
The more general matrix model ($p\neq0$) is obviously not necessary to match
the asymptotic behavior of the density of states of susy Yang-Mills integrals.
Note that free relativistic particle motion in $AdS$ space leads again to the
same polynomials \cite{Navarro:1994nu}. One can then use $AdS$/CFT ideas to
further establish the conformal symmetries of the model. Indeed, the
Klein-Gordon operator in $AdS$ is the quantum realization of the Casimir
operator of the Lie algebra of $SL(2,%
\mathbb{R}
)$ \cite{AdlGuer}. Thus, the matrix model turns out to be constructed from the
wavefunctions in the discrete series of the $SL(2,%
\mathbb{R}
)$ unitary irreducible representations.

An important property of a random matrix model with the distribution $\left(
\ref{weight}\right)  $ is that while it has very different global properties
(density of states), in comparison with strongly confining matrix models like
Gaussian or polynomial models, it turns out to have the same local properties
(correlation functions). Indeed, the correlation functions are in the Gaussian
universality class \cite{Mehta}. This can be intuitively understood from the
particular form of the confining potential, which is $V(x)=\left(
N+\alpha\right)  \log\left(  1+x^{2}\right)  $ in the Hermitian case. In
ordinary matrix models, it is well-known that one needs a potential with at
least linear confinement \cite{Mehta}. In the present case, although we have a
very weak potential (only a finite number of moments of the weight function
exists for finite $N$), we do have a linear growth, but given by the
$N-$dependent part of the potential, that makes the potential strongly
confining in the limit $N\rightarrow\infty.$ Thus, the same mechanism that
leads to the $N$-independent power-law tail of the density of eigenvalues is
responsible for the Gaussian universality of the correlation functions. To
summarize, the weight function changes with the dimension $N$ of the matrix
model and then the following two properties go hand in hand:

\begin{enumerate}
\item The tails of the density of states are power-law tails whose decay is
independent of the dimension $N$ of the matrix model.

\item The correlation functions of the model are in the Gaussian universality
class in the $N\rightarrow\infty$ limit, in spite of the very different
behavior of the density of states.
\end{enumerate}

This last result is formally proved for the Cauchy case ($\alpha=0$) by
mapping the model -with a one-dimensional inverse stereographic projection-
into the circular random matrix ensemble \cite{Hua} (see also \cite{Brow}),
which is known to possess correlation functions in the Gaussian universality
class when $N\rightarrow\infty$ \cite{Mehta}$.$ The cases $\alpha
=1/2,5/2,13/2$ discussed here follow from this result, as the potential has
the same confining properties for large $N$ (and it is more confining for
finite $N$).

Therefore, if it is possible to qualitatively describe the behavior of the YM
integrals with the corresponding one matrix models (e.g. the susy case with
the $SL(2,%
\mathbb{R}
)$ model discussed here and the bosonic case with the Wigner-Dyson paradigm
\cite{Mehta}), then one would expect that the correlation functions of
Yang-Mills matrix models exhibit the same behavior in the large $N$ limit,
regardless of the presence of supersymmetry, in spite of their very different
behavior for the density of states.

\begin{acknowledgments}
The author is grateful to Karl Landsteiner for a useful discussion
\end{acknowledgments}

\end{document}